\documentclass[twocolumn,trackchanges]{aastex631}
\pdfoutput=1
\usepackage[T1]{fontenc}
\usepackage{fontawesome}
\usepackage{color}
\usepackage{amsmath}
\usepackage{natbib}
\usepackage{ctable}
\usepackage{bm}
\usepackage[normalem]{ulem} 
\usepackage{xspace}
\usepackage{paralist}
\usepackage{fontawesome}

\let\oldbibliography\thebibliography 
\renewcommand{\thebibliography}[1]{%
  \oldbibliography{#1}%
  \setlength{\itemsep}{0pt}%
  \setlength{\parsep}{0pt}%
  \setlength{\parskip}{0pt}%
  \setlength{\bibsep}{0ex}
  \raggedright
}
\newcommand{\given}{\,|\,}

\newcommand{\lcdm}{$\Lambda$CDM}

\newcommand{\bfi}[1]{\textbf{\textit{#1}}}

\newcommand{\eg}{\emph{e.g.}}

\defcitealias{simbig_letter}{H22a}
\defcitealias{simbig_mock}{H23}

\let\oldAA\AA
\renewcommand{\AA}{\text{\normalfont\oldAA}}

\newcommand{\btheta}{\boldsymbol{\theta}}
\newcommand{\bphi}{\boldsymbol{\phi}}

\newcommand{\simbig}{{\sc SimBIG}}

\newcommand{\bitem}{\begin{itemize}}
\newcommand{\eitem}{\end{itemize}}
\newcommand{\beq}{\begin{equation}}
\newcommand{\eeq}{\end{equation}}

\definecolor{orange}{rgb}{1,0.5,0}

\begin{document} \sloppy\sloppypar\frenchspacing 

\title{{\sc SimBIG}: The First Cosmological Constraints from Non-Gaussian and Non-Linear Galaxy Clustering}
\newcounter{affilcounter}
\author[0000-0003-1197-0902]{ChangHoon Hahn}
\altaffiliation{changhoon.hahn@princeton.edu.com}
\affil{Department of Astrophysical Sciences, Princeton University, Princeton NJ 08544, USA} 

\author{Pablo Lemos}
\affil{Department of Physics, Universit\'{e} de Montr\'{e}al, Montr\'{e}al, 1375 Avenue Th\'{e}r\`{e}se-Lavoie-Roux, QC H2V 0B3, Canada}
\affil{Mila - Quebec Artificial Intelligence Institute, Montr\'{e}al, 6666 Rue Saint-Urbain, QC H2S 3H1, Canada}
\affil{Center for Computational Mathematics, Flatiron Institute, 162 5th Avenue, New York, NY 10010, USA}

\author{Liam Parker}
\affil{Department of Astrophysical Sciences, Princeton University, Princeton NJ 08544, USA} 

\author[0000-0003-0055-0953]{Bruno R\'egaldo-Saint Blancard}
\affil{Center for Computational Mathematics, Flatiron Institute, 162 5th Avenue, New York, NY 10010, USA}

\author{Michael Eickenberg}
\affil{Center for Computational Mathematics, Flatiron Institute, 162 5th Avenue, New York, NY 10010, USA}

\author{Shirley Ho}
\affil{Center for Computational Astrophysics, Flatiron Institute, 162 5th Avenue, New York, NY 10010, USA}

\author{Jiamin Hou}
\affil{Department of Astronomy, University of Florida, 211 Bryant Space Science Center, Gainesville, FL 32611, USA}
\affil{Max-Planck-Institut f\"ur Extraterrestrische Physik, Postfach 1312, Giessenbachstrasse 1, 85748 Garching bei M\"unchen, Germany}

\author[0000-0002-0637-8042]{Elena Massara}
\affil{Waterloo Centre for Astrophysics, University of Waterloo, 200 University Ave W, Waterloo, ON N2L 3G1, Canada}
\affil{Department of Physics and Astronomy, University of Waterloo, 200 University Ave W, Waterloo, ON N2L 3G1, Canada}

\author{Chirag Modi}
\affil{Center for Computational Mathematics, Flatiron Institute, 162 5th Avenue, New York, NY 10010, USA}
\affil{Center for Computational Astrophysics, Flatiron Institute, 162 5th Avenue, New York, NY 10010, USA}

\author[0000-0001-8841-9989]{Azadeh Moradinezhad Dizgah}
\affil{D\'epartement de Physique Th\'eorique, Universit\'e de Gen\`eve, 24 quai Ernest Ansermet, 1211 Gen\`eve 4, Switzerland}

\author{David Spergel}
\affil{Center for Computational Astrophysics, Flatiron Institute, 162 5th Avenue, New York, NY 10010, USA}
\affil{Department of Astrophysical Sciences, Princeton University, Princeton NJ 08544, USA}

\begin{abstract}
    The 3D distribution of galaxies encodes detailed cosmological information on the 
    expansion and growth history of the Universe.
    We present the first cosmological constraints that exploit non-Gaussian cosmological
    information on non-linear scales from galaxy clustering, inaccessible with current 
    standard analyses.
    We analyze a subset of the BOSS galaxy survey using \simbig, a new framework for 
    cosmological inference that leverages high-fidelity simulations and deep generative 
    models. 
    We use two clustering statistics beyond the standard power spectrum: the bispectrum and a convolutional neural network based summary of the galaxy field.
    We infer constraints on $\Lambda$CDM parameters, $\Omega_b$, $h$, $n_s$, 
    $\Omega_m$, and $\sigma_8$, that are 1.6, 1.5, 1.7, 1.2, and 2.3$\times$ tighter than
    power spectrum analyses.
    With this increased precision, we derive constraints on the Hubble constant, $H_0$, and 
    $S_8 = \sigma_8 \sqrt{\Omega_m/0.3}$ that are competitive with other cosmological 
    probes, even with a sample that only spans 10\% of the full BOSS volume. 
    Our $H_0$ constraints, imposing the Big Bang Nucleosynthesis prior on the baryon density, are consistent with the early time constraints from the cosmic microwave background (CMB). Meanwhile, our $S_8$ constraints are consistent with weak lensing experiments and similarly lie below CMB constraints. Lastly, we present forecasts to show that future work extending \simbig~to upcoming 
    spectroscopic galaxy surveys (DESI, PFS, {\em Euclid}) will produce leading $H_0$ and
    $S_8$ constraints that bridge the gap between early and late time measurements and shed 
    light on current cosmic tensions.
\end{abstract} 
\keywords{cosmological parameters from LSS --- Machine learning --- cosmological simulations --- galaxy surveys}

\section{Introduction} \label{sec:intro} 
The standard cosmological model (\lcdm) has been remarkably successful 
at describing a wide range of cosmological observations, including 
the CMB~\citep{page2003first, bennett2013nine, collaboration2020planck, aiola2020atacama, dutcher2021measurements}, 
the large-scale structure~\citep[LSS;][]{bernardeau2002, alam2017}, 
distances to Type Ia supernovae~\citep[SN-Ia;][]{perlmutter1999, riess1998, scolnic2018, brout2022}, 
and the observed abundance of light
elements~\citep{schramm1998, steigman2007, iocco2009, cyburt2016}. 
Despite its success, \lcdm~has recently come under scrutiny from 
`cosmic tensions', whose statistical significance has continued to persist 
with the latest observations~\citep[see][for a review]{abdalla2022}.

The most statistically significant of the tensions is the ``Hubble 
tension''. 
It refers to the disagreement between the late-time measurements
of the Hubble constant, $H_0$, with early-time measurements inferred from 
CMB assuming 
\lcdm~\citep[see][for recent reviews]{freedman2021, abdalla2022, kamionkowski2022}. 
With the latest observations~\citep{riess2022, collaboration2020planck}, the disagreement
has a statistical significance of >5$\sigma$. 
Meanwhile, there is also an ``$S_8$ tension'', which refers to the 
disagreement among measurements of the amplitude of the matter 
clustering, parameterized as $S_8 = \sigma_8 \sqrt{\Omega_m/0.3}$.
Weak lensing analyses that probe LSS at $z\sim0.5$ 
have systematically found $S_8$ values $\sim$2$\sigma$ lower 
than expected from the best-fit CMB \lcdm~cosmology~\citep{troxel2018, asgari2021, amon2022, secco2022, dalal2023, sugiyama2023}.  
Their $S_8$ values are also lower than the recent constraints 
from CMB lensing, which probes LSS at higher redshifts, 
$z = 0.5-5$~\citep{madhavacheril2023}.
The $H_0$ and $S_8$ tension between late- and early-time constraints, 
have led a number of theoretical works to explore 
alternatives to \lcdm~\citep[\eg][]{meerburg2014, chudaykin2018, divalentino2020, abellan2022, kamionkowski2022}. 


The 3D distribution of galaxies encodes cosmological information that 
addresses both tensions.
We can compare the angular and physical scales of features in the 
galaxy distribution, \eg~Baryon Acoustic 
Oscillations~\citep[BAO;][]{eisenstein1998a,
eisenstein2005, cole2005}, to compute physical 
distances and extract the Hubble constant~\citep[see][for a recent review]{ivanov2023}.
We can also measure the growth of structure from redshift-space distortions in the 
galaxy distribution to infer $S_8$ parameter. 
With these aims, spectroscopic galaxy surveys of the next decade, the 
Dark Energy Spectroscopic Instrument~\citep[DESI;][]{desicollaboration2016, desicollaboration2016a, abareshi2022}, 
Subaru Prime Focus Spectrograph~\citep[PFS;][]{takada2014, tamura2016}, 
the ESA {\em Euclid} satellite mission~\citep{laureijs2011}, and the
Nancy Grace Roman Space Telescope~\citep[Roman;][]{spergel2015, wang2022a}, 
will probe galaxies over unprecedented cosmic volumes.
They will span >10 Gyrs of cosmic history, $0 < z \lesssim 3$, and 
precisely constrain $H_0$ and $S_8$ from the early- to late-times of the Universe. 

To extract the cosmological information in galaxy distributions, 
current analyses primarily use the power spectrum, which captures all the information
in a Gaussian random field, as the summary statistic of galaxy
clustering~\citep[\emph{e.g.}][]{alam2017, beutler2017, ivanov2020, chen2022, kobayashi2022}. 
Furthermore, they model the galaxy clustering using the perturbation theory (PT) of LSS
and, as a result, are limited to large scales where deviations from linear 
theory are small ($k_{\rm max} \lesssim 0.2\,h/{\rm Mpc}$). 
However, recent studies have now established that there is additional 
cosmological information beyond these regimes. 
Analyses of the bispectrum, the first higher-order statistic, have shown 
that there is significant non-Gaussian 
information~\citep{gil-marin2017, philcox2022, ivanov2023a}. 
Forecasts in \cite{hahn2020} and \cite{hahn2021a} have further shown that cosmological
constraints improve by a factor of $\sim$2 by analyzing the bispectrum down to 
non-linear scales ($k_{\rm max} = 0.5\,h/{\rm Mpc}$).
Forecasts of other clustering statistics beyond the power spectrum find 
consistent improvements~\citep{massara2020, massara2022, wang2022, hou2022, eickenberg2022}.

Despite its promise, non-Gaussian cosmological information on non-linear 
scales cannot be robustly extracted using standard approaches. 
PT struggles to accurately model galaxy clustering beyond weakly non-linear scales.
It also cannot model many of the newly proposed summary 
statistics~\citep[\emph{e.g.}][]{banerjee2021, eickenberg2022, valogiannis2022, naidoo2022}. 
Observational systematics that significantly impact observed clustering are also a 
major challenge~\citep[\emph{e.g.}][]{ross2012, ross2017}.
Fiber collisions, for example, prevent galaxy surveys that use fiber-fed 
spectrographs (\emph{e.g.} DESI, PFS) from successfully measuring redshifts 
of more than one galaxy within some angular scale.
Even for the power spectrum, they cause significant bias on weakly non-linear scales,
$k > 0.1\,h/{\rm Mpc}$~\citep{guo2012}.
While proposed corrections may be sufficient for the power 
spectrum~\citep{hahn2017a, pinol2017, bianchi2018, smith2019},
no correction has yet been designed for statistics beyond the power spectrum. 

With the SIMulation-Based Inference of Galaxies (\simbig), we go beyond the 
state-of-the-art galaxy clustering analyses. 
As we present in the first papers of the series (\citealt{simbig_letter} 
and \citealt{simbig_mock}\footnote{hereafter \citetalias{simbig_letter} and \citetalias{simbig_mock}}), \simbig~is a forward modeling 
framework that uses simulation-based inference\footnote{also known as ``likelihood-free inference'' (LFI) or ``implicit likelihood inference'' (ILI)}~\citep[SBI; see][for a review]{cranmer2020}
with deep generative models from machine learning
to perform highly efficient cosmological inference. 
\simbig~leverages high-fidelity $N$-body simulations that accurately model non-linear 
galaxy clustering and includes observational effects.
With this approach, \citetalias{simbig_letter} analyzed the power spectrum multipoles,
$P_\ell(k)$, of galaxies\footnote{We used 109,636 galaxies in the Southern Galactic 
Cap (SGC) of the BOSS CMASS sample within $0.45 < z < 0.6$ and over
$\sim$3,600 ${\rm deg}^{2}$, which corresponds to roughly $\sim$10\% of the full 
BOSS volume.} 
in the Sloan Digital Sky Survey-III Baryon Oscillation Spectroscopic Survey~\citep[BOSS;][]{eisenstein2011, dawson2013} to demonstrate that we can robustly analyze the 
power spectrum down to non-linear scales.
In this work, we go further and present the cosmological constraints that exploit 
{\em non-Gaussian} information and non-linear scales. 
We use two summary statistics of galaxy clustering beyond $P_\ell$: the galaxy 
bispectrum and a field-level summary based on convolutional neural networks.
We explore the cosmological implications of our constraints and 
address the $H_0$ and $S_8$ tensions. 

We begin by briefly summarizing our methods in Section~\ref{sec:simbig}.
We present and discuss our cosmological results in Section~\ref{sec:results} and compare
them to the literature. 
Lastly, we discuss future steps and prospects of extending \simbig~to 
upcoming galaxy surveys in Section~\ref{sec:future}.
\section{Methods} \label{sec:simbig}
\simbig~uses SBI to infer posteriors of $\Lambda$CDM cosmological parameters with  only a forward model that can generate mock observations, \emph{i.e.} the 3D galaxy spatial distribution. 
In this section, we briefly describe the forward model, the SBI 
methodology, the considered summaries of the galaxy distribution, and the validation 
of our analyses. 

\subsection{Forward Model} \label{sec:fm} 
The {\sc Simbig} forward model constructs simulated galaxy catalogs
from {\sc Quijote}  $N$-body simulations~\citep{villaescusa-navarro2020} 
run at different cosmologies in a Latin-hypercube configuration.  
Each simulation has a volume of $1~(h^{-1}{\rm Gpc})^3$ and is constructed using
$1024^3$ cold dark matter particles initialized at $z=127$ and gravitationally 
evolved until $z=0.5$.
These high-resolution $N$-body simulations accurately model the
clustering of matter down to non-linear scales beyond $k > 0.5\,h/{\rm Mpc}$.
From the dark matter particles, halos are identified using the 
{\sc Rockstar} halo finder~\citep{behroozi2013a}, which accurately determines the 
location of halos and resolve their substructure~\citep{knebe2011}. 
Afterwards, the halos are populated using the halo occupation 
distribution~\citep[HOD; \eg][]{berlind2002} framework, which provides a flexible 
statistical prescription for determining the number of galaxies as well as their 
positions and velocities within  halos.  
{\sc SimBIG} uses a state-of-the art 9-parameter HOD model that supplements the 
standard~\cite{zheng2007} model with assembly, concentration, and velocity biases for 
even more flexibility. 
From the HOD galaxy catalog, {\sc SimBIG} adds full BOSS survey realism by applying
the survey geometry and observational systematics. 
The forward-modeled catalogs have the same redshift range and 
the angular footprint of our observed sample. 
In summary, the \simbig~forward model aims to generate mock 
galaxy catalogs that are statistically indistinguishable from the 
observations.
For further details, we refer readers to \citetalias{simbig_letter} and \citetalias{simbig_mock}. 

\subsection{Simulation-Based Inference} \label{sec:sbi} 
From the 20,000 forward-modeled galaxy catalogs, we use the {\sc SimBIG} 
SBI framework to infer posterior distributions of cosmological 
parameters, $\btheta$, for given a summary statistic, $\bfi{x}$,
of the observations: $p(\btheta\given\bfi{x})$. 
The SBI in \simbig~is based on neural density estimation (NDE) from 
deep generative models and uses ``normalizing flow''~\citep{tabak2010, tabak2013}. 
It enables cosmological inference with a limited number of simulated forward models
so that we can extract cosmological information on small, non-linear, scales and 
using any statistic. 

Among various normalizing flow architectures, {\sc SimBIG} uses Masked Autoregressive
Flow~\citep[MAF;][]{papamakarios2017} and Neural Spline Flow~\citep[NSF;][]{durkan2019}
models\footnote{We use the normalizing flow 
implementation in $\mathtt{sbi}$ Python package~\citep{greenberg2019,
tejero-cantero2020}.}.
We train a flow that best approximates the posterior, 
$q_{\bphi}(\btheta\given\bfi{x}) \approx p(\btheta\given\bfi{x})$,
by minimizing the KL divergence between 
$p(\btheta, \bfi{x}) = p(\btheta\given\bfi{x}) p(\bfi{x})$ and
$q_{\bphi}(\btheta\given\bfi{x}) p(\bfi{x})$. 
In practice, we maximize the total log-likelihood 
$\sum_i \log q_{\bphi}(\btheta_i\given \bfi{x}_i)$ over the 
training set. 
We determine the training and architecture of our flow through experimentation,
tailored to each summary statistic. 

The prior of our posterior estimate is set by the parameter distribution of our 
training set, which corresponds to a uniform prior over the cosmological parameters,
$\{\Omega_m, \Omega_b, h, n_s, \sigma_8\}$, with ranges that extend well beyond the 
{\em Planck} posterior.
For the HOD parameters, we use the conservative priors centered around previous HOD 
analyses of the BOSS CMASS sample. 

\subsection{Summary Statistics Beyond $P_\ell$} \label{sec:stats}
We use two summary statistics of galaxy clustering beyond $P_\ell$: 
the bispectrum ($B_0$) and a field-level summary using
convolutional neural networks (CNNs). 
The bispectrum is the first higher-order statistic that measures the 
excess probability of finding galaxies in different triangle 
configurations over a random distribution. 
For a near-Gaussian galaxy distribution, the bispectrum extracts nearly 
all of its cosmological information~\citep[\emph{e.g.}][]{fry1994, matarrese1997, scoccimarro2000}.
We also analyze the galaxy distribution at the field-level, using  
CNNs to perform massive data compression and extract maximally relevant 
features.
By learning the relevant features, we aim to extract even more 
cosmological information than conventional summary statistics.  
We provide additional details on these summary statistics in 
Appendix~\ref{sec:stats}. 
We also refer readers to \cite{simbig_bk} and \cite{simbig_cnn} 
for further details.

\subsection{Posterior Validation} \label{sec:valid}
For each summary statistic, we validate the trained $q_\phi$ in two steps. 
First, we perform a ``NDE accuracy test'' on a held-out subset of the training 
data, either using simulation-based calibration~\citep{talts2020} or 
a ``distance to random point'' test~\citep{lemos2023}. 
With this test, we check whether $q_\phi$ accurately estimates
the posterior throughout the entire prior range of the cosmological parameters. 
Second, we assess the robustness of $q_\phi$ by applying it to the 
\simbig~``mock challenge'' simulations (\citetalias{simbig_mock}). 
These simulations include three different sets of test simulations evaluated at some
fiducial cosmology: 
500 constructed using the same
forward model as the training data, 
500 constructed using a different
halo finder and HOD model, and 
1,000 constructed using a different $N$-body simulation and halo finder. 
We apply $q_\phi$ to the simulations and compare the resulting posteriors to their 
true parameter values. 
Since two of the test sets are constructed using different forward models than 
the training simulations, if the $q_\phi$ posteriors derived from them are
statistically consistent with the true values, we conclude that the $q_\phi$ 
is sufficiently robust.

\begin{figure*}
    \centering
    \includegraphics[width=17.8cm]{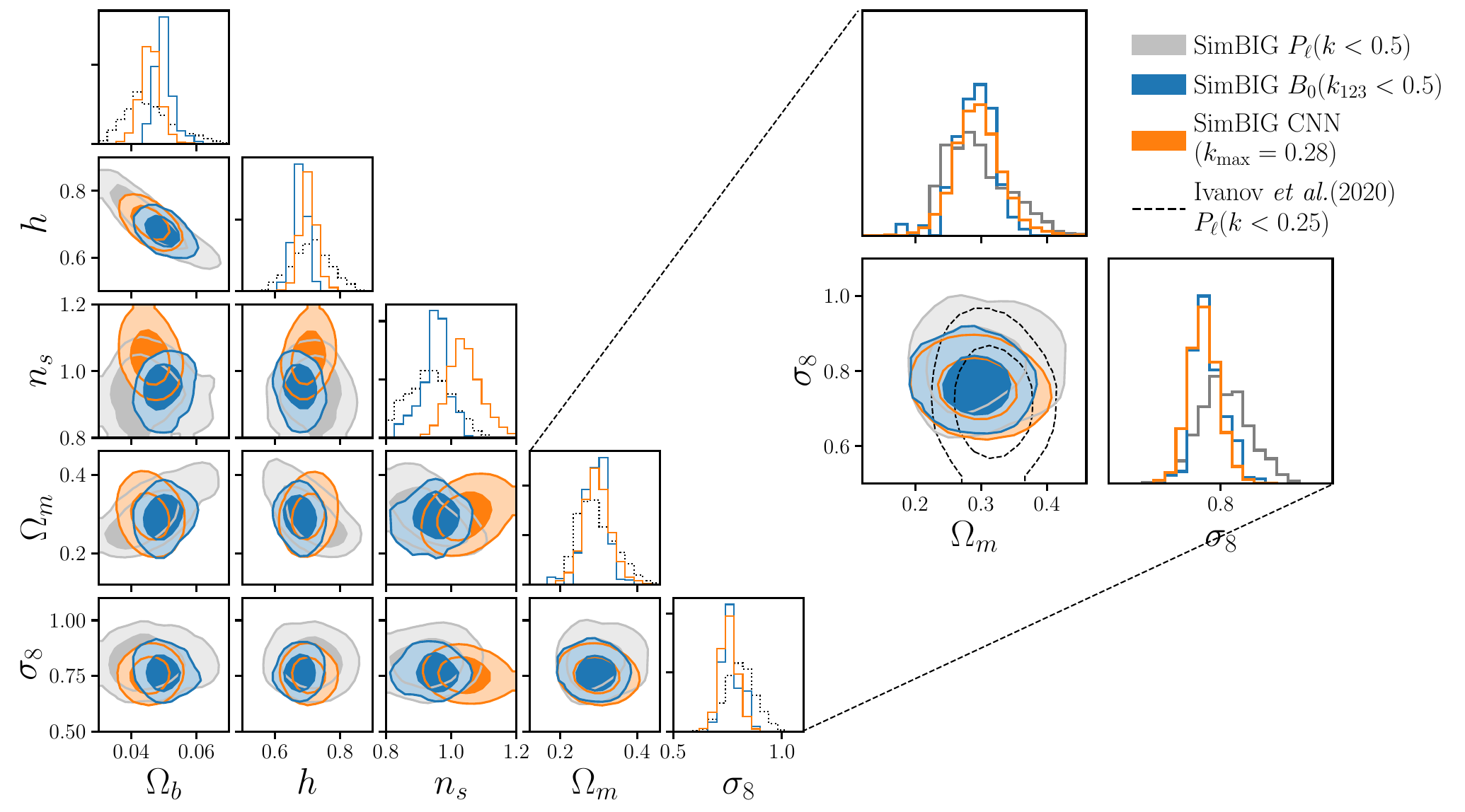}
        \caption{
        {\em Left}: 
        Posteriors of cosmological parameters inferred from $B_0$ (blue) and CNN (orange) 
        using \simbig. 
        All posteriors include a $\omega_b$ prior from BBN studies. 
        The contours mark the 68 and 95 percentiles.
        For comparison, we include the posterior from the \simbig~$P_\ell$ analysis 
        (gray). 
        {\em Right}: 
        We focus on the posteriors of $\Omega_m$ and $\sigma_8$, the
        parameters that can be most significantly constrained by galaxy clustering.  
        We include the posterior from the \cite{ivanov2020} PT-based 
        $P_\ell(k < 0.25\,h/{\rm Mpc})$ analysis for reference (black dashed). 
        The \simbig~$B_0$ and CNN constraints are significantly tighter
        yet consistent with $P_\ell$ constraints. 
        }
    \label{fig:post}
    \centering
\end{figure*}

\section{Results} \label{sec:results}
We present the posteriors of the $\Lambda$CDM cosmological parameters for the $B_0$ 
(blue) and CNN (orange) \simbig~analyses in Figure~\ref{fig:post}.
In the right panels, we focus on the constraints on the growth of structure 
parameters: $\Omega_m$ and $\sigma_8$.
For comparison, we include the posteriors from the \citetalias{simbig_letter} 
\simbig~(gray) and \cite{ivanov2020} (black dashed) $P_\ell$ analyses. 
The \cite{ivanov2020} constraints are inferred from roughly the same galaxy sample and
is based on PT. 
In each of the contours, we mark the 68 and 95 percentiles of the posteriors.
For all posteriors, we impose a Gaussian prior on 
$\omega_b = \Omega_b h^2 = 0.02268\pm0.00038$ from
BBN studies~\citep{aver2015, cooke2018, schoneberg2019}, typically included in clustering analyses~\citep[\emph{e.g.}][]{ivanov2020, philcox2022}.
We list the marginalized parameter constraints in Table~\ref{tab:post}.

Focusing first on $B_0(k_{123} < 0.5\,h/{\rm Mpc})$, we find that our \lcdm~constraints 
are significantly tighter yet consistent with $P_\ell$ constraints. 
Compared to the \simbig~$P_\ell(k < 0.5\,h/{\rm Mpc})$ constraints, the 
\simbig~$B_0$ constraints are 
2.78, 3.17, 1.91, 1.77, and 1.53$\times$ tighter for $\Omega_b$,  $h$, $n_s$, 
$\Omega_m$, and $\sigma_8$.
Compare to the \cite{ivanov2020} $P_\ell(k < 0.25\,h/{\rm Mpc})$ constraints, 
the $B_0$ constraints 
are 1.59, 1.51, 1.68, 1.17, and 2.04$\times$ tighter for $\Omega_b$, $h$, $n_s$, 
$\Omega_m$, and $\sigma_8$, respectively. 
The improvement on $\sigma_8$ is larger because \cite{ivanov2020} does not include 
the non-linear scales (0.25 < $k$ < 0.5\,$h/{\rm Mpc}$) included in the
\simbig~$P_\ell$ analyses.

Similarly, our constraints from the \simbig~CNN analysis are significantly tighter 
yet consistent with the $P_\ell$ constraints. 
For $\Omega_b$, $h$, $n_s$, $\Omega_m$, and $\sigma_8$, the CNN constraints are
2.51, 2.51,1.46, 1.42, and 1.75$\times$ tighter than the \simbig~$P_\ell$ constraints.
In comparison to the PT $P_\ell$ analysis, the $\Omega_b$, $h$, $n_s$, and $\sigma_8$
constraints are 1.44, 1.20, 1.28, and 2.33$\times$ tighter while the
$\Omega_m$ constraint is slightly (1.07$\times$) broader. 

Overall, the CNN and $B_0$ constraints are in excellent agreement. 
There is a slight offset in $n_s$; however it is not statistically significant. 
The CNN constraints are slightly broader overall than $B_0$ except for $\sigma_8$, 
even though the CNN aims to maximally extract the cosmological information in 
the galaxy field. 
This is in part due to the fact that, in practice, we limit the CNN's constraining
power of the CNN to ensure robustness and generalizability 
(Appendix~\ref{sec:stats}). 

\begin{table*} 
    \centering
    \begin{tabular}{l|ccccccc} 
        \hline
         & $\Omega_m$ & $\sigma_8$ & $\Omega_b$ & $h$ & $n_s$ & $S_8$ & $H_0$\\[3pt]
        \hline
        $P_\ell(k<0.5)$     &  $0.287^{+0.059}_{-0.038}$ &
        $0.808^{+0.068}_{-0.066}$ & $0.044^{+0.009}_{-0.006}$ &  
        $0.713^{+0.056}_{-0.059}$ &  $0.935^{+0.070}_{-0.073}$ &  
        $0.797^{+0.107}_{-0.093}$ & 
        $71.255^{+5.615}_{-5.921}$ 
                             \\[3pt]
        $B_0(k_{123}<0.5)$  & 
        $0.297^{+0.020}_{-0.035}$ & 
        $0.763^{+0.054}_{-0.033}$ & 
        $0.050^{+0.003}_{-0.002}$ & 
        $0.676^{+0.018}_{-0.018}$ & 
        $0.952^{+0.035}_{-0.040}$ & 
        $0.757^{+0.048}_{-0.053}$ & 
        $67.556^{+1.801}_{-1.843}$ \\[3pt]
        CNN                 & 
        $0.295^{+0.036}_{-0.033}$ & 
        $0.753^{+0.040}_{-0.036}$ & 
        $0.046^{+0.003}_{-0.003}$ & 
        $0.702^{+0.024}_{-0.022}$ & 
        $1.039^{+0.053}_{-0.044}$ & 
        $0.746^{+0.055}_{-0.051}$ &
        $70.199^{+2.387}_{-2.208}$ \\[3pt]
        \hline            
    \end{tabular} \label{tab:post}
    \caption{Posteriors of $\Lambda$CDM cosmological parameters
    inferred from the power spectrum multipoles, the bispectrum
    monopole, and the CNN using \simbig. 
    We present the median and 68 percentile uncertainties of the 
    parameters. 
    All posteriors include a $\omega_b$ prior from BBN.} 
\end{table*}

Overall, the \simbig~analyses beyond $P_\ell$ produce significantly 
tighter cosmological constraints than $P_\ell$ analyses.
Taking the best constraints from the $B_0$ and CNN analyses, we improve 
$\Omega_b$, $h$, $n_s$, $\Omega_m$, and $\sigma_8$ by 1.59, 1.51, 1.68, 
1.17, and 2.33$\times$ over the PT $P_\ell$ analysis. 
This firmly shows that there is substantial non-Gaussian cosmological information 
in galaxy clustering.
Furthermore, our posteriors are in excellent agreement with previous 
galaxy clustering analyses that use an entirely different approach. 
This consistency serves as additional corroboration that 
\simbig~provides a robust framework for analyzing clustering 
beyond $P_\ell$.

\begin{figure}
\begin{center}
    \includegraphics[width=0.4\textwidth]{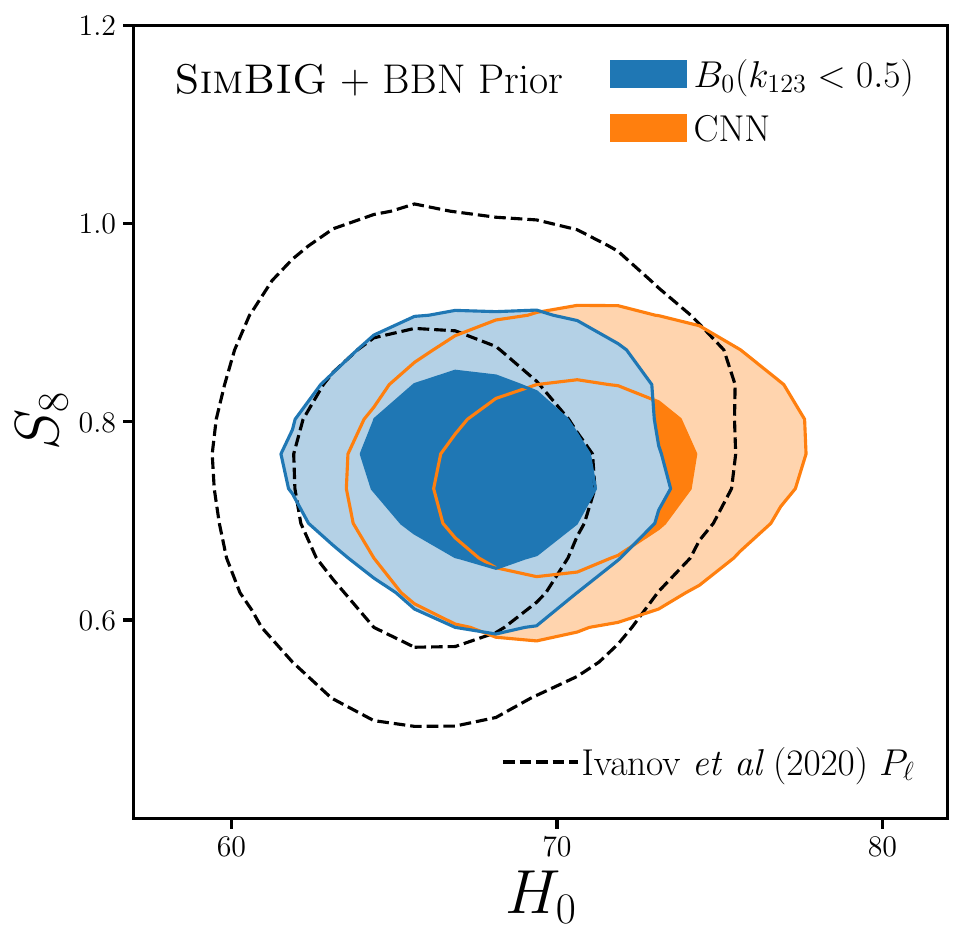}
    \caption{\label{fig:S8H0}
        $S_8$ and $H_0$ posteriors for the \simbig~$B_0$ (blue) and CNN (orange) analyses. 
        We include the posterior from the \cite{ivanov2020} $P_\ell$ analyses 
        of CMASS-SGC for comparison (black dashed). 
        All posteriors include a $\omega_b$ prior from BBN studies. 
        With the $B_0$ and CNN, we infer consistent posteriors on $S_8$ and $H_0$ that 
        are $\sim$1.9 and 1.5$\times$ tighter than the PT $P_\ell$ analyses. 
        Our tighter constraints enable us to inform the $S_8$ and $H_0$ ``tensions'' 
        found between cosmological probes in the early versus late Universe. 
    }
\end{center}
\end{figure}

\begin{figure*}
\begin{center}
    \includegraphics[width=\textwidth]{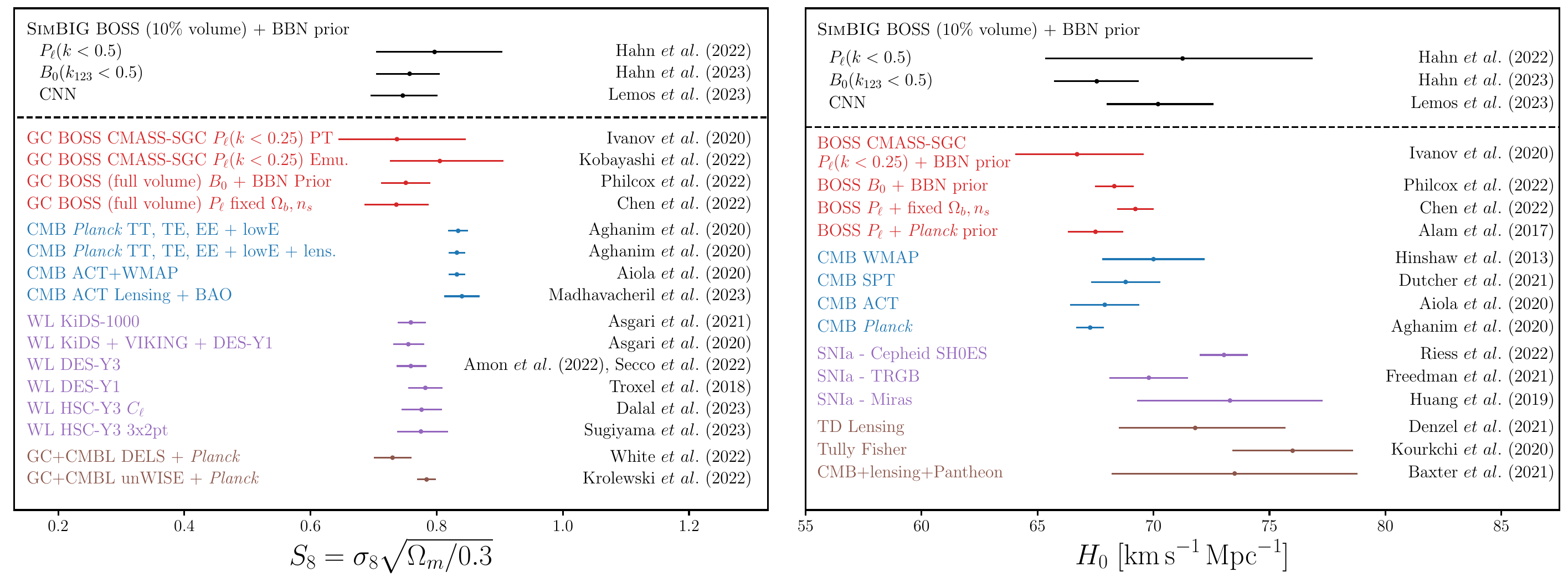}
    \caption{\label{fig:S8H0_comp}
        {\em Left}: Comparison of the \simbig~$S_8$ constraints (black)
        to existing bounds in the literature.
        We include constraints from galaxy clustering (red), CMB (blue),   
        weak lensing (purple), and multi-probe (brown) analyses. 
        Despite only using 10\% of the full BOSS volume, we derive $S_8$ 
        constraints comparable to full BOSS $P_\ell$ analysis.
        Our $S_8$ constraints are in good agreement with weak lensing experiments 
        and, thus, lie below the CMB constraints. 
        {\em Right}: 
        Comparison of the \simbig~$H_0$ constraints (black) to the literature.
        We include constraints from galaxy clustering (red), CMB (blue), 
        SN (purple), and other probes (brown).
        Our $H_0$ constraint from \simbig~$B_0$ is in excellent agreement with 
        {\em Planck}. 
        We infer a higher $H_0$ with the CNN that is consistent with both 
        {\em Planck} and SN-Ia constraints. 
    }
\end{center}
\end{figure*}

\subsection{Cosmic Tensions}
Our tighter constraints on cosmological parameters enable us to also inform the  
$S_8$ and $H_0$ cosmic tensions. 
In Figure~\ref{fig:S8H0}, we present the $S_8$ and $H_0$ posteriors from 
the \simbig~$B_0$ (blue) and  CNN (orange) analyses. 
For comparison, we also plot the posterior for the \cite{ivanov2020}~$P_\ell$ PT analysis (black
dashed).
We again include a $\omega_b$ prior from BBN studies.
The $S_8$ and $H_0$ constraints from $B_0$ and CNN are consistent 
with each other (Table~\ref{tab:post}). 
Compared to \cite{ivanov2020}, $B_0$ improves $S_8$ and $H_0$ constraints by 1.92 and 1.51$\times$ and the CNN improves the $S_8$ and $H_0$ by 1.83 and 1.20$\times$. 

Next, we compare our $S_8$ and $H_0$ posteriors to constraints in the literature 
from galaxy clustering and other cosmological probes (Figure~\ref{fig:S8H0_comp}).
The constraints from the literature are selected from \cite{abdalla2022}. 
In the left panel, we compare the \simbig~$S_8$ constraints (black) to constraints 
from galaxy clustering (red), CMB (blue), weak lensing (purple), and multi-probe 
(brown) analyses. 
From galaxy clustering, we include the \cite{ivanov2020} and 
\cite{kobayashi2022} $P_\ell$ analyses of the CMASS-SGC sample. 
These analyses use nearly the same observational sample so their constraints
are the most consistently comparable to the 
\simbig~constraints\footnote{The galaxy sample in our \simbig~analyses has a $\sim$30\% 
smaller footprint than the CMASS-SGC.}. 
We also include constraints from PT analyses of the full BOSS volume: \cite{philcox2022} 
and \cite{chen2022}.
\cite{philcox2022} analyze both $P_\ell$ and $B_0$ and includes a BBN prior;
\cite{chen2022} analyze the pre- and post-reconstruction $P_\ell$  with fixed
$\Omega_b$ and $n_s$.
Despite only analyzing 10\% of the total BOSS volume, our analyses 
produce $S_8$ constraints with comparable precision level to full volume analyses. 

For CMB, we include constraints from: 
{\em Planck} TT,TE,EE+lowE with and without CMB lensing~\citep{aghanim2020}; 
the~\cite{aiola2020} combined high-$\ell$ and low-$\ell$ measurements from the Atacama Cosmology  Telescope (ACT) and the Wilkinson Microwave Anisotropy Probe (WMAP); 
and the ACT CMB lensing constraint combined with BAO~\citep{madhavacheril2023}.
For weak lensing (WL), we include: 
the cosmic shear analysis of Kilo Degree Survey~\citep[KiDS;][]{asgari2021};
the \cite{asgari2020} combined analysis of earlier KiDS data, VIKING data, and the Dark Energy Survey (DES) first year data;
the DES analyses of the first~\citep{troxel2018} and third year data~\citep{amon2022, secco2022};
and the Hyper Suprime Cam (HSC) cosmic shear analysis~\citep{dalal2023} and 3$\times$2pt 
analysis~\citep{sugiyama2023} of the third year data. 
Lastly, we include multi-probe analyses that combine galaxy clustering and the cross-correlation
with the {\em Planck} CMB lensing reconstruction~\citep{krolewski2021, white2022}. 

Next, in the right panel of Figure~\ref{fig:S8H0_comp}, we compare the 
\simbig~$H_0$ posteriors to constraints from a wide variety of cosmological probes: 
galaxy clustering (red), CMB (blue), SN-Ia (purple), and others
(brown).
From galaxy clustering, we include the \cite{ivanov2020} CMASS-SGC analysis. 
Again, this is the analysis most consistent with our analyses. 
We also include the \cite{philcox2021}, \cite{chen2022}, and \cite{alam2017} 
analyses of the full BOSS volume.
From CMB, we include constraints from WMAP~\citep{hinshaw2013}, 
{\em Planck}~\citep{aghanim2020}, ACT~\citep{aiola2020}, and the South 
Pole Telescope~\citep[SPT;][]{dutcher2021}.
We also include $H_0$ constraints from distance ladder methods: SN-Ia 
calibrated using Cepheids from the SH0ES collaboration~\citep{riess2022} and 
SN-Ia calibrated using the Tip of the Red Giant 
Branch~\citep[TRGB;][]{freedman2021}.
Lastly, we include constraints from time-delay strong lensing~\citep{denzel2021}, 
the Tully-Fisher relation~\citep{kourkchi2020}, and gravitational wave standard 
sirens~\citep{palmese2020}.

The \simbig~$P_\ell$ constraints are not precise enough to meaningfully inform the 
$S_8$ and $H_0$ tensions.
However, the tighter constraints from the \simbig~$B_0$ and CNN analyses provide
interesting insights into the tension. 
For $S_8$, the \simbig~$B_0$ and CNN constraints are in good agreement with the
weak lensing constraints and, thus, lower than the CMB constraints. 
Despite the differences, we emphasize that the \simbig~constraints are statistically 
consistent with both CMB and weak lensing. 
For $H_0$, the tighter constraints from \simbig~$B_0$ and CNN are in good agreement with 
constraints from CMB and LSS.
The $B_0$ constraint is in excellent agreement with {\em Planck} and, thus, is in 
tension with the SH0ES measurement. 
The CNN constraint is notably higher than the $B_0$ constraint and consequently 
reduces the tension with SH0ES. 
Nevertheless, it is statistically consistent with $B_0$ and {\em Planck}.

\subsection{Future Steps and Upcoming Surveys} \label{sec:future}
Despite only using 10\% of the BOSS volume, we derive $S_8$ and 
$H_0$ constraints that are competitive with other cosmological  
probes and galaxy clustering analyses  of the full BOSS volume. 
In short, our forward-modeling approach improves $S_8$ and $H_0$ constraints by
$\sim$2.0 and 1.5$\times$ over the standard PT $P_\ell$ analysis. 
The improvement on $S_8$ is equivalent to applying the standard PT 
$P_\ell$ analysis to a galaxy sample {\em four times} the cosmic 
volume. 

Yet, the improvement does not include {\em all} of the cosmological information that 
can be extracted using clustering statistics beyond $P_\ell$. 
We require \simbig~analyses to derive unbiased cosmological constraints for a suite 
of test simulations constructed with different forward models (Section~\ref{sec:valid}). 
To meet this requirement for the CNN analysis, we use dropout, regularization, and SWA 
to reduce overfitting (Appendix~\ref{sec:stats}).
These procedures discard a substantial amount of  cosmological information.

\begin{figure}
\begin{center}
    \includegraphics[width=0.4\textwidth]{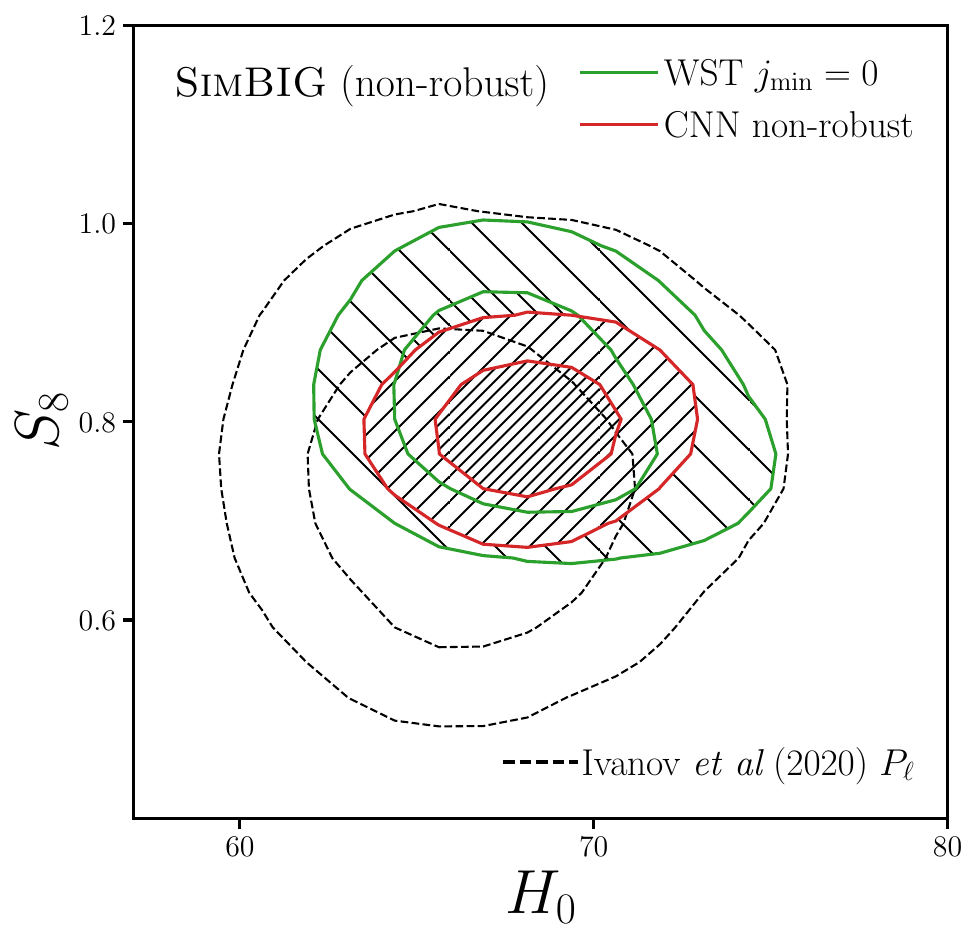}
    \caption{\label{fig:notrobust}
        The $H_0$ and $S_8$ posterior for the CNN (red hatched) with {\em relaxed
        robustness requirements}.
        We also include the posterior from the \simbig~WST analysis (green hatched).
        The posteriors are {\em not} robust; however, they illustrate the potential of
        the WST and CNN at capturing even more cosmological information than the
        posteriors presented above.
        The WST is capable of deriving constraints comparable to
        $B_0$ while the non-robust CNN can double the precision of the robust 
        CNN constraints.
        In future work, we aim to extract some of this additional constraining 
        power using \simbig~with a more flexible and generalizable forward model 
        and with summary statistics more robust to model misspecifications.
    }
\end{center}
\end{figure}

Solely for reference, in Figure~\ref{fig:notrobust} we present the $\Omega_m$ and
$\sigma_8$ posterior for the  
CNN (red hatched) {\em if we relax our robustness requirement}.
This is {\em not} a robust posterior; however, it illustrates that the CNN is capable of 
capturing significantly more cosmological information than the posteriors presented in 
this work. 
The non-robust CNN has >2$\times$ the precision of the robust CNN. 
In Figure~\ref{fig:notrobust}, we also include posteriors derived 
from another summary statistic: the wavelet scattering transform (WST). 
The WST is set of descriptive statistics particularly suited 
for characterizing non-Gaussian fields. 
It is motivated by the structure of CNNs while being completely 
parameter-free. 
We exclude the WST analysis in our main comparison due to concerns about its robustness;
however, the WST is capable of deriving constraints comparable to the $B_0$ analysis. 
For further details on the WST analysis, see \cite{simbig_wst}.

The full constraining power of neither the CNN nor the WST can currently be robustly
exploited. 
This is because these statistics identify cosmological imprints that are specific 
to our forward model. 
One way to robustly exploit these statistics is to use a more flexible and generalizable
forward model that can describe the WST and CNN across the different test simulations.  
Alternatively, we can also construct clustering statistics more robust to
model misspecification~\citep[\eg~][]{huang2023}. 
Such an approach would learn the imprints of cosmological parameters while 
ignoring any model-specific features. 
In future work we will explore both approaches. 

\begin{figure*}
\begin{center}
    \includegraphics[width=0.6\textwidth]{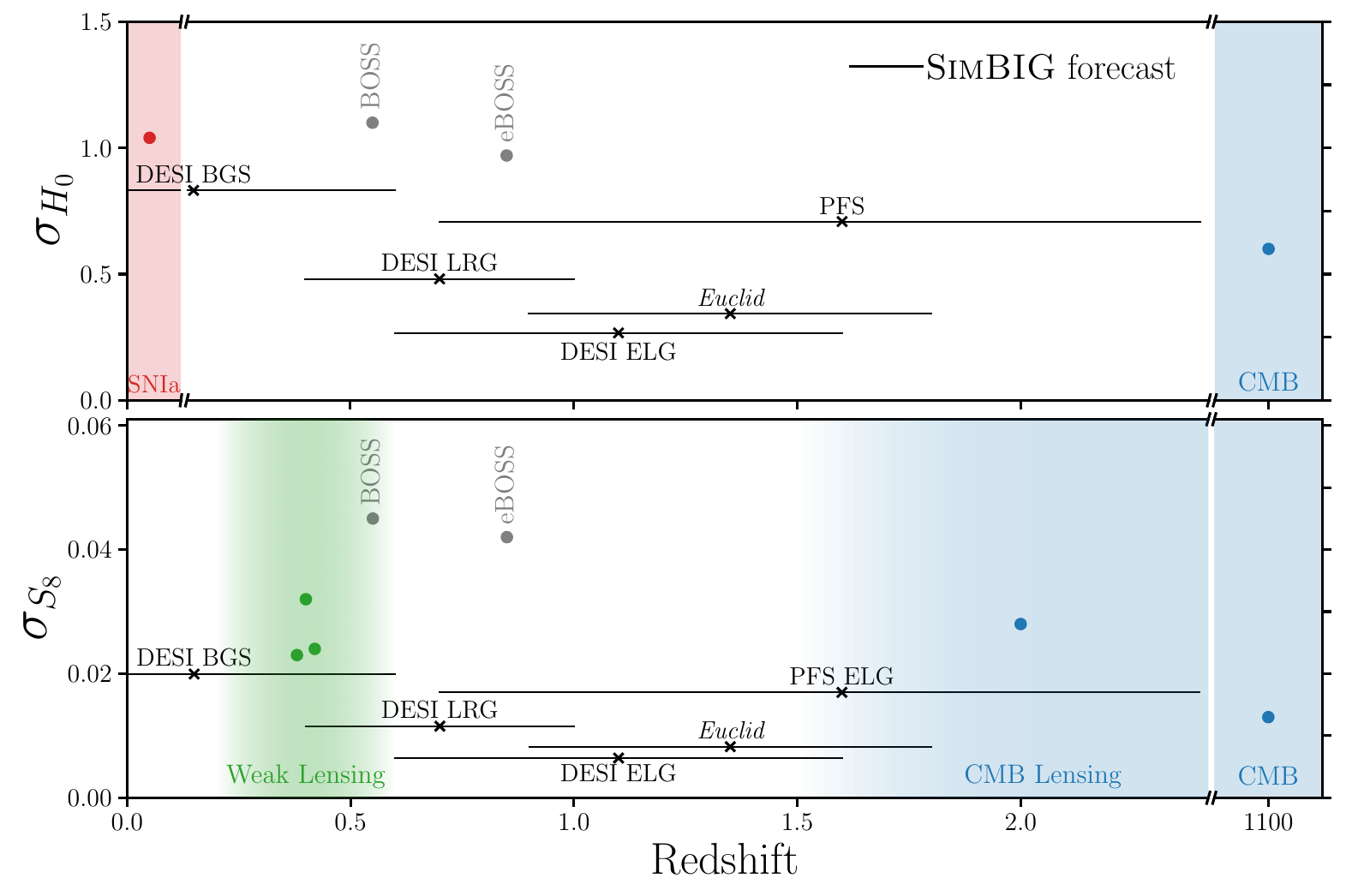}
    \caption{\label{fig:forecast}
        Expected 1$\sigma$ precision level of $H_0$ ($\sigma_{H_0}$; top) and $S_8$ 
        ($\sigma_{S_8}$; bottom) constraints from applying \simbig~to upcoming galaxy surveys, 
        DESI, PFS, {\em Euclid} (black). 
        For DESI, we present \simbig~forecasts for the different galaxy samples 
        individually.
        The width of $\sigma_{H_0}$ and $\sigma_{S_8}$ marks the redshift range of each 
        sample.
        We include constraints from SN-Ia (red), CMB and CMB lensing (blue), 
        weak lensing (green),
        and previous BOSS/eBOSS $P_\ell$ analyses (gray), for comparison.
        Future \simbig analyses of DESI, PFS, and {\em Euclid} will have the precision 
        and redshift range to provide key input into the $H_0$ and $S_8$ tensions
        and potentially reveal new physics beyond the standard $\Lambda$CDM model.
    }
\end{center}
\end{figure*}

In future work, we will also extend \simbig~to the next generation spectroscopic 
galaxy surveys. 
DESI will probe 14,000 deg$^2$ across $0 < z < 2.1$ with >40 million
galaxies~\citep{chaussidon2023, hahn2022c, raichoor2023, zhou2023}. 
DESI has recently completed two out of its five years of observations. 
Meanwhile, the PFS Cosmology Survey will probe 1,200 deg$^2$ across $0.6 < z < 2.4$ with 
4 million emission-line galaxies and is set to begin observations next year. 
Lastly, the {\em Euclid} spectroscopic survey~\citep{euclid} will measure
the redshift of about 30 million H$\alpha$ emitters across 
15,000 ${\rm deg}^2$ over $0.9 < z < 1.8$. 
It was recently launched in July 2023.

These surveys will probe enormous cosmic volumes in epochs without precise constraints 
on $H_0$ or $S_8$ from other cosmological probes. 
We highlight this in Figure~\ref{fig:forecast}, where we present the 
forecasted 1$\sigma$ precision levels of $H_0$ (top) and $S_8$ (bottom) 
constraints from applying \simbig~to DESI, PFS, and {\em Euclid} (black). 
These forecasts are derived by scaling the \simbig~constraints by $(V_{\rm survey} / V_{\rm CMASS-SGC})^{0.5}$.
For $H_0$, upcoming surveys will help bridge the gap 
between the late-time measurements from SN-Ia (SH0ES; red)
and early-time measurements from the CMB 
({\em Planck}; blue).
$H_0$ constraints from other probes, \eg~time delay lensing, have significantly 
larger uncertainties and lie well outside the range of the figure.
Similarly, for $S_8$, DESI, PFS, and {\em Euclid} will bridge the gap between $S_8$
constraints from  weak lensing surveys that probe low redshifts and 
from CMB lensing and CMB at high redshifts.
The shaded regions represent the lensing efficiency kernel for weak 
lensing experiments~\citep[0.1 < $z$ < 0.7;][]{amon2022} and the mass-map weights 
for CMB lensing~\citep[0.5 < $z$ < 5.0;][]{madhavacheril2023}.

\simbig~can produce leading constraints on both $S_8$ and $H_0$ by more fully 
extracting the cosmological information of upcoming galaxy surveys. 
We note that the \simbig~forecasts are conservative estimates that 
do not account for the fact that the DESI, PFS, and {\em Euclid} 
galaxy samples will have a higher number density than CMASS.
Given their higher number density, we expect even larger gains from 
analyzing clustering beyond $P_\ell$.
With their precision levels and the extensive redshift range of
DESI, PFS, and {\em Euclid}, future
\simbig~analyses will provide critical input into the cosmic tensions and 
potentially reveal new physics beyond the standard $\Lambda$CDM model.

\section{Summary} \label{sec:summary}
We present a comparison of \simbig~cosmological constraints from 
analyzing galaxy clustering beyond the power spectrum: the bispectrum 
and a field-level summary statistic based on CNNs.  
\simbig~provides a forward-modeling framework that uses 
simulation-based inference with normalizing flows to perform 
highly efficient cosmological inference using high-fidelity 
$N$-body simulations  
(\citetalias{simbig_letter}, \citetalias{simbig_mock}). 
With \simbig, we extract non-Gaussian and non-linear cosmological information, currently 
inaccessible with standard PT analyses. 

The \simbig~$B_0$ and CNN analyses tightly constrain all 
$\Lambda$CDM parameters and, compared to a standard PT analysis of $P_\ell$~\citep{ivanov2020}, improve
$\Omega_b$, $h$, $n_s$, $\Omega_m$, and $\sigma_8$ constraints by 1.6, 1.5, 
1.7, 1.2, and 2.3$\times$. 
With these tighter constraints, the \simbig~analyses can inform recent ``tensions'' 
between the early and late time measurements of $S_8$ and $H_0$. 
Even with a subset set of the SDSS-III BOSS CMASS-SGC sample that corresponds to 
only <10\% of the full volume, we derive $S_8$ constraints competitive with PT $P_\ell$
analyses of the full BOSS volume. 
Our $S_8$ constraint is statistically consistent with both CMB and weak lensing experiments. 
We also infer competitive $H_0$ constraints 
that are consistent with early universe constraints from CMB and other 
LSS analyses and in tension with late-time measurements. 

In future work, we will focus on extending \simbig~in two main avenues. 
First, we will increase the flexibility of our forward model in order
to improve its robustness and generalizability. 
With a more robust forward model, we will use more constraining statistics that can 
improve cosmological constraints even further (Figure~\ref{fig:notrobust}). 
We will also explore constructing new summary statistics that are
simultaneously more sensitive to cosmological imprints and more robust
to model misspecifications.

In addition, we will extend \simbig~to upcoming spectroscopic 
galaxy surveys: DESI, PFS, and {\em Euclid}. 
These surveys will probe unprecedented volumes with >70 million galaxies. 
\simbig~can produce leading constraints on both $S_8$ and
$H_0$ across $0 < z < 2.4$ by more fully extracting the cosmological 
information from these surveys (Figure~\ref{fig:forecast}).   
The \simbig~analyses of these surveys will bridge the gap between the early- and 
late-time $H_0$ and $S_8$ constraints to provide key input into the tensions and 
potentially reveal deviations from the standard $\Lambda$CDM cosmology model.

\section*{Acknowledgements}
It's a pleasure to thank Peter Melchior, Uro\u{s} Seljak, Benjamin D. Wandelt, 
and the members of the Simons Collaboration on Learning the
Universe\footnote{\url{https://www.learning-the-universe.org/}} for valuable discussions. 
We also thank Mikhail M. Ivanov and Yosuke Kobayashi for providing us with 
the posteriors used for comparison. 
This work was supported by the AI Accelerator program of the Schmidt Futures Foundation. 
JH has received funding from the European Union's Horizon 2020 research and innovation program under the Marie Sk\l{}odowska-Curie grant agreement No 101025187. AMD acknowledges funding from Tomalla Foundation for Research in Gravity. 
The work reported on in this paper was substantially performed using the Princeton Research Computing resources at Princeton University, which is a consortium of groups led by the Princeton Institute for Computational Science and Engineering (PICSciE) and Office of Information Technology’s
Research Computing.

\appendix
\section{Supplementary Information}

\subsection{Summary Statistics} \label{sec:stats}

{\bf The bispectrum}: $B(k_1, k_2, k_3)$, is the three-point
correlation function in  Fourier space and measures the excess
probability of different triangle configurations $(k_1, k_2, k_3)$
over a random distribution. 
In this work, we use the monopole of the bispectrum: 
$B_0(k_1, k_2, k_3)$
To measure $B_0$, we use the redshift-space bispectrum estimator by~\cite{scoccimarro2015}. 
The estimator accounts for the survey geometry using a random
catalog that has the same radial and angular selection functions
as the observed catalog but with >4 million objects.
For each galaxy, we include the \cite{feldman1994} weights.
For the observed galaxy sample, we also include angular systematic weights to 
account for stellar density and seeing conditions as well as redshift failure weights. 
We do not include weights for fiber collisions, since this effect is included in the 
\simbig~forward model.  
We measure $B_0$ in triangle configurations defined by $k_1, k_2, k_3$ bins of 
width $\Delta k = 0.0105\,h/{\rm Mpc}$.
In practice, we use the reduced bispectrum, $Q_0(k_1, k_2, k_3)$, which normalizes the 
bispectrum to reduce the dynamic range of $B_0$. 
For $k_{\rm max} = 0.5\,h/{\rm Mpc}$, $Q_0$ has 10,052 total 
triangle configurations. 


{\bf CNN}: Convolutional Neural Networks (CNNs) can be thought of 
as flexible, hierarchical, non-linear functions that can be 
optimized to extract relevant features from input images 
or, in this case, 3D galaxy fields. 
In this work, we use a 3D CNN to perform a step of massive data compression from the galaxy sample directly to the cosmological parameters.
First, the galaxy distributions are meshed to 
$64 \times 64 \times 128$ voxels using cloud-in-cell mass assignment.
The voxels have dimensions $\sim [11, 11, 11]~h/\textrm{Mpc}$, which 
imposes a scale cut of $k < k_{\rm max} = 0.28~ h/\text{Mpc}$. 
We then train the CNN to accurately compress the galaxy field by 
minimizing the mean-squared-error between the network-predicted
cosmological parameters and the true parameter values of 
\simbig~training set using stochastic gradient descent (SGD).
In addition, we also perform weight marginalization on the network 
using Stochastic Weight Averaging~\citep[SWA;][]{maddox2019simple, wilson2020bayesian},
to improve the generalizability of our CNN~\citep{lemos2023robust}.

\bibliography{simbig} 
\end{document}